\documentclass[a4paper, oneside]{article}
\usepackage[english]{babel}
\usepackage[T1]{fontenc} 
\usepackage{lmodern}
\usepackage[cp1251]{inputenc} 
\usepackage[tbtags]{amsmath}
\usepackage{amsfonts,amssymb,mathrsfs,amscd}
\usepackage{amsthm}
\usepackage{graphicx}
\usepackage{cite}
\usepackage{hyperref}
\hypersetup{
	colorlinks=true,
	linkcolor=blue,
	filecolor=blue,
	urlcolor=blue,
} 
    
\newcommand{\Tr}{{\rm Tr}}
\begin{document}
	
\title{\bf\Large{Minimal Time Generation of Density Matrices for a~Two-Level Quantum System Driven by Coherent and Incoherent Controls}}
 
\author{\normalsize {\bf Oleg~V.~Morzhin}\footnote{E-mail: {\tt morzhin.oleg@yandex.ru}}~$^{,1}$ \quad and \quad 
	{\bf Alexander~N.~Pechen}\footnote{E-mail: {\tt apechen@gmail.com} (corresponding author)}~$^{,1,2}$ \vspace{0.2cm} \\
	\small $^1$ Steklov Mathematical Institute of Russian Academy of Sciences,\\
	\small Department of Mathematical Methods for Quantum Technologies, \\
	\small 8 Gubkina Str., Moscow, 119991, Russia; \\
	\small $^2$ National University of Science and Technology ``MISiS'',\\
	\small 6 Leninskiy prospekt, Moscow, 119991, Russia} 

\date{ }

\maketitle

\makeatletter
\renewcommand{\@makefnmark}{}
\makeatother 
  
\begin{abstract}
The article considers a two-level open quantum system whose dynamics is driven 
by a combination of coherent and incoherent controls. Coherent control enters 
into the Hamiltonian part of the dynamics whereas incoherent control enters 
into the dissipative part. The goal is to find controls which move the system 
from an initial density matrix to a given target density matrix as fast as 
possible. To achieve this goal, we reformulate the optimal control problem 
in terms of controlled evolution in the Bloch ball and then apply Pontryagin
maximum principle and gradient projection method to numerically find minimal time 
and optimal coherent and incoherent controls. General method is provided 
and several examples of initial and target states are explicitly considered.
	
\vspace{0.3cm} {\bf Key words}: Quantum control, open quantum system, coherent control, 
incoherent control.
\end{abstract}

\tableofcontents
 
\section{Introduction}  
Control of quantum systems of atomic and molecular scale is an important branch of modern
science with existing and prospective applications in physics, chemistry and quantum
technology~\cite{RiceBook2000, BrumerBook2003, TannorBook2007, 
	FradkovBook2007, WisemanBook2010, Petersen2010, ZagoskinBook2011, Brif_Chakrabarti_Rabitz_article_2010, 
	Glaser2015Report, CPKoch_2016_OpenQS, Borzi_book_2017, Lyakhov_Pechen_Lee_2018, Amosov_Mokeev_2018, Avanesov_Kronberg_Pechen_2018}. Two general types of quantum control exist. 
Coherent control drives essentially the Hamiltonian aspects of the dynamics, and is typically realized by a shaped
laser pulse~\cite{RiceBook2000, BrumerBook2003, TannorBook2007, 
	FradkovBook2007, Brif_Chakrabarti_Rabitz_article_2010}. Incoherent control drives 
non-Hamiltonian, i.e., dissipative aspects of the dynamics, and can be realized 
by reservoir engineering~\cite{Pechen_Rabitz_2006}, 
measurement apparatus~\cite{WisemanBook2010,Pechen_Trushechkin_2015}, etc. 

Control of quantum systems driven by a combination of coherent and incoherent controls
was considered in~\cite{Pechen_Rabitz_2006, Pechen_Rabitz_2014}. Then it was 
shown that for any initial and target states of a general $n$-level quantum system 
there exist a combination of coherent and incoherent controls which move the initial 
state arbitrarily close to the target state asymptotically as final time
$T\to\infty$~\cite{Pechen_PhysRevA_2011}. Therefore the quantum system under coherent 
and incoherent controls is asymptotically controllable in the space of all density matrices. 
However, method of~\cite{Pechen_PhysRevA_2011} does not guarantee that this state-to-state 
transfer is as fast as possible. This motivates the problem of finding a way 
to steer the initial state to a final state in a minimal possible time, which 
we study below for a two-level system.  

Time-optimal control problems were considered for open and closed two-level systems,
for example, in~\cite{Lapert_2010,Boscain_Gronberg_Long_Rabitz_2014, Albertini_DAlessandro_2016}. 
Optimal control at the quantum speed limit for a two-level Landau-
Zener system is analyzed~\cite{Caneva2009}. Manipulation of states of a degenerate quantum system is
considered~\cite{Volovich_Kozyrev_2016}. In this article we consider minimal time steering of an initial density matrix
of a two-level system into a target density matrix by coherent and incoherent controls. We
reformulate the control problem as evolution of a real vector in the Bloch ball, then apply
Pontryagin maximum principle~\cite{Pontryagin_et_al_book_1962} and a version of gradient projection method 
(GPM)~\cite{Nikolskii_2007, Demyanov_Rubinov_book_1970, Bertsekas_book_2016}. Using GPM in the functional 
space of piecewise continuous control functions, we compute sequential improvements 
of controls for various $\rho_0$ and $\rho_{\rm target}$.

\section{Formulation of the Problem} 

Most general state of a two-level system is described by a density matrix, i.e., by a~$(2 \times 2)$ Hermitian matrix $\rho(t) \in \mathbb{C}^{2 \times 2}$ which is positive, $\rho(t) \geq 0$, and has unit trace, $\Tr \rho(t) = 1$. Evolution of the density matrix is described by the master equation 
(see \cite{Pechen_Rabitz_2006})
\begin{eqnarray}
	\dfrac{d \rho(t)}{dt} &=& 
	-\dfrac{i}{\hbar} \Big[ \widehat{\bf H}_0 + 
	\widehat{\bf V} v(t), \rho(t) \Big] + 
	\gamma D(\rho(t), n(t)), \qquad \rho(0) = \rho_0. \label{f1}
\end{eqnarray}
Operators $\widehat{\bf H}_0$ 
and $\widehat{\bf V}$ are Hermitian, and $\widehat{\bf H}_0$ has two different eigenvalues. Without loss of generality we consider
\[
\widehat{\bf H}_0 = \hbar \omega \begin{pmatrix}
0 & 0 \\
0 & 1  
\end{pmatrix},\qquad \widehat{\bf V} = \mu \begin{pmatrix}
0 & 1 \\
1 & 0
\end{pmatrix},
\]
where $\omega > 0$, $\mu \in \mathbb{R}$, $\mu \neq 0$.
Dissipative superoperator describes interaction between the system and its environment and has the form
\begin{eqnarray}
	D(\rho(t), n(t)) &=& n(t) \Big( \sigma^+ \rho(t) \sigma^- + \sigma^- \rho(t) \sigma^+   
	- \dfrac{1}{2} \Big\{ \sigma^- \sigma^+ + \sigma^+ \sigma^-, \rho(t) \Big\} \Big) + \nonumber \\
	&+&\Big( \sigma^+ \rho(t) \sigma^- - \dfrac{1}{2} \left\{ \sigma^- \sigma^+, \rho(t)\right\}\Big). \label{f2}
\end{eqnarray}
The parameter $\gamma > 0$ determines strength of interaction with the environment. Matrices $\sigma^\pm$ are
\[
\sigma^- = \begin{pmatrix}
0 & 0 \\ 1 & 0
\end{pmatrix},\qquad
\sigma^+ = \begin{pmatrix}
0 & 1 \\ 0 & 0
\end{pmatrix}.
\] 
We use the notations for commutator $[A, B] = AB - BA$ 
and anti-commutator $\{A, B\} = AB + BA$ of two operators $A$ and $B$.
Function $v = v(t)$, $t \in [0, T]$ represents a coherent control 
(e.g., shaped laser field), where $T$ is the final time. Function $n(t)$, $t \in [0, T]$ represents incoherent control (for example, non-equilibrium spectral density or temperature of the environment). The incoherent control by its physical meaning is a non-negative function. Consider controls $v$ and $n$ as piecewise continuous functions bounded as $v_{\min} \leq v(t) \leq v_{\max}$, $0 \leq n(t) \leq n_{\max}$. Thus
\begin{equation} 
	\begin{array}{c}
		u = (v, n) \in \mathcal{U} = PC([0,T]; Q), \qquad
		Q = [v_{\min}, v_{\max}] \times [0, n_{\max}] \subset \mathbb{R}^2.
	\end{array} \label{f3}
\end{equation}
Consider for the system (\ref{f1}) --- (\ref{f3}) the following terminal
constraint:
\begin{equation}
	\rho(T) = \rho_{\rm target} \label{f4}
\end{equation} 
where $\rho_{\rm target}$ is some given target density matrix. The reachable set $\mathcal{R}(T, \rho_0, \mathcal{U})$ for the system (\ref{f1}) --- (\ref{f3}) is the set of all states $\rho(t)$ which can be 
obtained from $\rho_0$ by controls from $\mathcal{U}$
to the time $T$. It can happen that for small enough $T$ 
the corresponding reachable set $\mathcal{R}(T, \rho_0, \mathcal{U})$ does not
contain the target state $\rho_{\rm target}$. 

The problem of moving the system (\ref{f1}), (\ref{f3}) from a given initial
density matrix $\rho_0$ to a given target density matrix $\rho_{\rm target}$ during as small as possible time $T$ can be formulated as minimization of the objective
\begin{equation}
	J(u,T) = T \to \min \label{f5}
\end{equation}
subject to constraint (\ref{f4}). In other words, the goal is to find   
\begin{gather*}
	\overline{T} = \min\left\{T> 0~|~ \rho(T) = \rho_{\rm target} \right\} 
\end{gather*}
and the corresponding control $\overline{u} \in \mathcal{U}$. 

\section{Evolution in the Bloch Ball} 

In this section we reformulate the original control problem as controlled evolution in the Bloch ball.

Consider the representation of density matrix (e.g., \cite{Holevo_book_De_Gruyter_2012})
\begin{equation}
	\rho = \dfrac{1}{2} \left( \sigma_0 + \sum\limits_{j=1}^3 x_j \sigma_j \right) = 
	\dfrac{1}{2} \begin{pmatrix}
		1 + x_3 & x_1 - i x_2 \\ x_1 + i x_2 & 1 - x_3
	\end{pmatrix}, \label{f6}
\end{equation}
where matrices $\sigma_0 = \mathbb{I}_2$, $\sigma_1 = \begin{pmatrix}
0 & 1 \\ 1 & 0
\end{pmatrix}$,
$\sigma_2 = \begin{pmatrix}
0 & -i \\ i & 0
\end{pmatrix}$,
$\sigma_3 = \begin{pmatrix}
1 & 0 \\ 0 & -1
\end{pmatrix}$ form the Pauli basis. Vector $x = (x_1, x_2, x_3) \in \mathbb{R}^3$ 
satisfies the condition $\| x \|^2 \leq 1$. For pure quantum states
vector $x$ satisfies the condition 
$\| x \|^2 = 1$ for each $t$, i.e. the points $x$ evolve on the Bloch sphere.
If $\| x\|^2 < 1$, then $x$ represent a mixed quantum state and such $x$ are located 
in the inner part of the Bloch ball. The point in the origin represents 
the completely mixed state.

Using (\ref{f6}), we rewrite the system (\ref{f1}), (\ref{f2}) as
\begin{eqnarray}
	\dfrac{dx_1}{dt} &=& -\dfrac{\gamma}{2} x_1 + \omega x_2 - \gamma x_1 n, \qquad x_1(0) = x_{1,0}, \label{f7} \\
	\dfrac{dx_2}{dt} &=& -\omega x_1 - \dfrac{\gamma}{2} x_2 - 2\kappa x_3 v - \gamma x_2 n, 
	\qquad x_2(0) = x_{2,0}, \label{f8} \\
	\dfrac{dx_3}{dt} &=& 2 \kappa x_2 v - \gamma x_3 + \gamma - 2 \gamma x_3 n, 
	\qquad x_3(0) = x_{3,0}, \label{f9}
\end{eqnarray}
where $\kappa = \mu/\hbar$. The terminal constraint (\ref{f4}) takes the form
\begin{equation}
	x_i(T) = x_{i,\rm target}. \label{f10}
\end{equation}
The values $x_{i,0}$ and $x_{i,\rm target}$ are calculated for the given matrices
$\rho_0$ and $\rho_{\rm target}$ as $x_{i,0} = \Tr \rho_0 \sigma_i$ and 
$x_{i, \rm target} = \Tr \rho_{\rm target} \sigma_i$.

\section{Optimization Method} 

\subsection{Reducing to a Sequence of Fixed-Time Optimal Control Problems}

We apply for solving the optimal control problem (\ref{f3}), (\ref{f5}), (\ref{f7}) --- (\ref{f10})
the following approach. Consider a series of optimal control problems $P_j$, $j = 1, 2, \dots K$, 
where $j$th problem has no terminal constraint and is considered 
with some final time $T = T_j \in \left\{ T_1, T_2, \dots, T_K \right\}$. 
Cost criterion for each optimal control problem $P_j$ is 
\begin{equation}
	J_j(u) = \| x(T_j) - x_{\rm target} \|^2 \to \min. \label{f11}
\end{equation}
Thus, instead of the problem (\ref{f3}), (\ref{f5}), (\ref{f7}) --- (\ref{f10})
we consider a series of the problems (\ref{f3}), (\ref{f7}) --- (\ref{f9}), (\ref{f11}) 
for $j = 1, 2, \dots, K$. The goal is to obtain the minimal possible $T_j$ 
for which $J_j = 0$. Setting some sufficiently large $T_1$, we solve 
the sequence of optimal control problems until we can move
the system from the initial state $x_0$ to the target state $x_{\rm target}$. 

The constraint $\| x(t)\|^2 \leq 1$ 
is satisfied automatically by the evolution equation and there is no need to use a special method for taking into account this constraint. 

\subsection{Solving a Fixed-Time Optimal Control Problem}

This subsection considers GPM for solving a particular optimal control problem
of the type (\ref{f3}), (\ref{f7}) --- (\ref{f9}), (\ref{f11}). We omit the index $j$ in the final time $T_j$ for shorten the notation. 

We apply the Pontryagin maximum principle \cite{Pontryagin_et_al_book_1962} 
which uses the Pontryagin function and the conjugate variables. In this case, the Pontryagin function is 
\begin{eqnarray*}
	H(p,x,u) = \mathcal{K}_v(p,x) v + \mathcal{K}_n(p,x) n + \widetilde{H}(p,x),
	\label{Pontryagin_function_f1}
\end{eqnarray*} 
where $p \in \mathbb{R}^3$, the switching functions
\begin{eqnarray*}
	\mathcal{K}_v(p,x) &=& \dfrac{\partial H}{\partial v} = 
	2 \kappa \left(p_3 x_2 - p_2 x_3 \right), \label{Pontryagin_function_f2} \\
	\mathcal{K}_n(p,x) &=& \dfrac{\partial H}{\partial n} = 
	-\gamma \left( p_1 x_1 + p_2 x_2 + 2 p_3 x_3 \right), \label{Pontryagin_function_f3}
\end{eqnarray*} 
and
\begin{eqnarray*}
	\widetilde{H}(p,x) &=& p_1 \Big(-\dfrac{\gamma}{2} x_1 + \omega x_2 \Big)  
	+ p_2 \Big(-\omega x_1 - \dfrac{\gamma}{2} x_2 \Big)  
	+ p_3 \Big(\gamma - \gamma x_3 \Big). \label{Pontryagin_function_f4} 
\end{eqnarray*} 
The conjugate system is 
\begin{eqnarray}
	\dfrac{dp_1}{dt} &=& \dfrac{\gamma}{2} p_1 + \gamma p_1 n + \omega p_2, \label{f12} \\
	\dfrac{dp_2}{dt} &=& - \omega p_1 + \dfrac{\gamma}{2} p_2 + \gamma p_2 n - 2 \kappa p_3 v, \label{f13} \\
	\dfrac{dp_3}{dt} &=& 2 \kappa p_2 v + \gamma p_3 + 2 \gamma p_3 n, \label{f14} \\
	\qquad p_i(T) &=& -2 \left(x_i(T) - x_{i, \rm target}\right), \qquad i = 1, 2, 3. \label{f15}
\end{eqnarray}  
The gradient of the cost 
functional $J$ at some control $u \in \mathcal{U}$ is the following:
\begin{eqnarray}
	\dfrac{\delta J}{\delta u(t)} &=& 
	-\dfrac{\partial H}{\partial u}\left(p(t), x(t), u(t) \right) =  -\left(\mathcal{K}_v\left(p(t), x(t) \right), 
	\mathcal{K}_n\left(p(t), x(t) \right) \right), \label{gradient_f} 
\end{eqnarray}
where $x$, $p$ are correspondingly the solutions of the systems (\ref{f7}) --- (\ref{f9})
and (\ref{f12}) --- (\ref{f15}) for the considered control $u$.  

Fix some value $\alpha > 0$ which defines the step of the method, and 
$0 < \varepsilon \ll 1$ which defines the stopping criterion (e.g., $\varepsilon = 10^{-9}$). 
At $k$th iteration, GPM is represented by the following operations:
\begin{enumerate}
	\item For the current admissible process $(u^{(k)}, x^{(k)})$ 
	compute the corresponding solution $p^{(k)}$ of the system 
	(\ref{f12}) --- (\ref{f15}).
	\item Compute the gradient (\ref{gradient_f})
	at the triple of the functions $u = u^{(k)}$, $x=x^{(k)}$, $p=p^{(k)}$. 
	\item Form the function 
	\begin{eqnarray}
		u^{(k)}(t; \alpha) = u^{(k)}(t) - \alpha \dfrac{\delta J}{\delta u(t)}\Big|_{u = u^{(k)}}. \label{PGM_f0}
	\end{eqnarray} 
	The components $v^{(k)}$ and $n^{(k)}$ of the function $u^{(k)}(t; \alpha)$ are:
	\begin{eqnarray}
		v^{(k)}(t; \alpha) &=& v^{(k)}(t) + 
		\alpha \mathcal{K}_v(p^{(k)}(t), x^{(k)}(t)), \label{PGM_f1} \\
		n^{(k)}(t; \alpha) &=& n^{(k)}(t) + 
		\alpha \mathcal{K}_n(p^{(k)}(t), x^{(k)}(t)). \label{PGM_f2}
	\end{eqnarray}
	Further, form the function
	\begin{gather}
		u^{(k)}_{\rm Pr}(t; \alpha) = {\rm Pr}_Q(u^{(k)}(t; \alpha)), \label{ff}
	\end{gather}
	where ${\rm Pr}_Q$ is the orthogonal projection which maps any point outside of $Q$ to a closest
	point in $Q$, and leaves unchanged points in $Q$ \cite{Bertsekas_book_2016}. Its explicit action on
	the vector $u^{(k)}(t; \alpha)$ is the following: 
	\begin{eqnarray}
		v^{(k)}_{\rm Pr}(t; \alpha) &=& \begin{cases}
			v^{\min}, & v^{(k)}(t; \alpha) < v^{\min},\\
			v^{(k)}(t; \alpha), & v^{\min} \leq v^{(k)}(t; \alpha) \leq v^{\max}, \\
			v^{\max}, & v^{(k)}(t; \alpha) > v^{\max},  
		\end{cases} \label{PGM_f3} \\
		n^{(k)}_{\rm Pr}(t; \alpha) &=& \begin{cases}
			0, & n^{(k)}(t; \alpha) < 0,\\
			n^{(k)}(t; \alpha), & 0 \leq n^{(k)}(t; \alpha) \leq n^{\max}, \\
			n^{\max}, & n^{(k)}(t; \alpha) > n^{\max}. 
		\end{cases} \label{PGM_f4}  	
	\end{eqnarray}  
	\item Form the control   
	$u^{(k)}(\cdot; \alpha, \beta) 
	= \left(v^{(k)}(\cdot; \alpha, \beta), 
	n^{(k)}(\cdot; \alpha, \beta)\right)$ which components are
	\begin{eqnarray}
		v^{(k)}(t; \alpha, \beta) &=& v^{(k)}(t) + 
		\beta \left(v^{(k)}_{\rm Pr}(t; \alpha) - v^{(k)}(t) \right), \label{PGM_f5} \\
		n^{(k)}(t; \alpha, \beta) &=& n^{(k)}(t) + 
		\beta \left(n^{(k)}_{\rm Pr}(t; \alpha) - n^{(k)}(t) \right), \label{PGM_f6}
	\end{eqnarray}
	where $\beta \in (0,1]$. 
	\item Compute the value 
	\begin{eqnarray}
		\beta^{(k)} &=& {\rm arg}\min\limits_{\beta \in (0,1]} f(\beta), \label{PGM_f7}
	\end{eqnarray}
	where $f(\beta)=  
	J\left( u^{(k)}(\cdot; \alpha, \beta) \right)$.
	This iteration step is the hardest because the problem (\ref{PGM_f7}) 
	requires global optimization and the function $f(\beta)$ is defined
	implicitly such that for each $\beta$ the value $f(\beta)$ is computed
	through solving the Cauchy problem (\ref{f7}) --- (\ref{f9}) with the
	corresponding $u(\cdot) = u^{(k)}(\cdot; \alpha, \beta)$.
	\item Construct the next approximation 
	\begin{eqnarray}
		u^{(k+1)}(t) &=& u^{(k)}_{\rm Pr}(t; \alpha, \beta^{(k)}), \label{PGM_f8}
	\end{eqnarray}
	the corresponding solution $x^{(k+1)}$ of the Cauchy problem (\ref{f7}) --- (\ref{f9}),  
	and compute the value $J(u^{(k+1)})$. If the inequality
	\begin{eqnarray}
		\left| J(u^{(k+1)}) - J(u^{(k)}) \right| < \varepsilon
		\label{PGM_f9}
	\end{eqnarray}
	is satisfied, then stop the iteration process; otherwise, 
	take $k:= k +1$ and go to the next iteration.
\end{enumerate}

In this article, we consider control of two-level quantum systems. For such systems $\rho(t)$ is $(2 \times 2)$ matrix which admits convenient parametrization~(\ref{f6}) by a vector in the Bloch ball. In the general case of an $n$-level quantum system, $\rho(t)$ is an $(n \times n)$ 
matrix for which there is no such a simple parametrization. The extension of our method to this general case is an important task for a future work, 
that may require other parametrizations for $\rho(t)$, as for example, parametrization considered in~\cite{Ilin_Shpagina_Uskov_Lychkovskiy_article_2018}. 

GPM is a first order (based on gradient of the cost functional) method which contains at each iteration the computationally hard step~(\ref{PGM_f7}) 
for finding the most suitable variation of the control $u^{(k)}$. For the considered control problem, it may be useful to adapt the 2nd order Krotov method (e.g.,~\cite{Sklarz_Tannor_article_2002}), which in contrast with GPM, being a method for nonlocal improvements, does not use computationally hard variation 
of the control~$u^{(k)}$. Both GPM and the Krotov methods can give sequential improvements in the functional space of controls and can take 
into account constraints on the control values. It may be also useful to exploit GRAPE (GRadient Ascent Pulse Engineering)~\cite{Khaneja_Reiss_Kehlet_SchulteHerbruggen_Glaser_2005} and CRAB (Chopped Random-Basis Quantum 
Optimization)~\cite{Caneva_Calarco_Montangero_2011} methods. Both GPM and GRAPE use gradient of the cost functional but in contrast 
to GPM, GRAPE operates in a finite-dimensional space of parameters describing piecewise-constant parametrization of control functions. 
CRAB also works in a finite-dimensional space of parameters. However, these parameters describe trigonometric parametrization of the control. 
This parametrization takes into account the wave nature of coherent control. For optimization of the parameters in CRAB, one can use zero-order 
methods such as the Nelder-Mead method.

\section{Numerical Results} 

Consider in the system (\ref{f7}) --- (\ref{f9}), 
$\omega = 1$, $\gamma = 2 \times 10^{-3}$, $\kappa =10^{-2}$, 
and in (\ref{f3}), $v_{\min} = -10$, $v_{\max} = 10$, and $n_{\max} = 1$. 
Set the parameter $\alpha = 10^3$. The value of  $\alpha$ is chosen sufficiently large to compensate small values 
of the switching functions.

\begin{figure}
	\centering
	\includegraphics[scale=0.37]{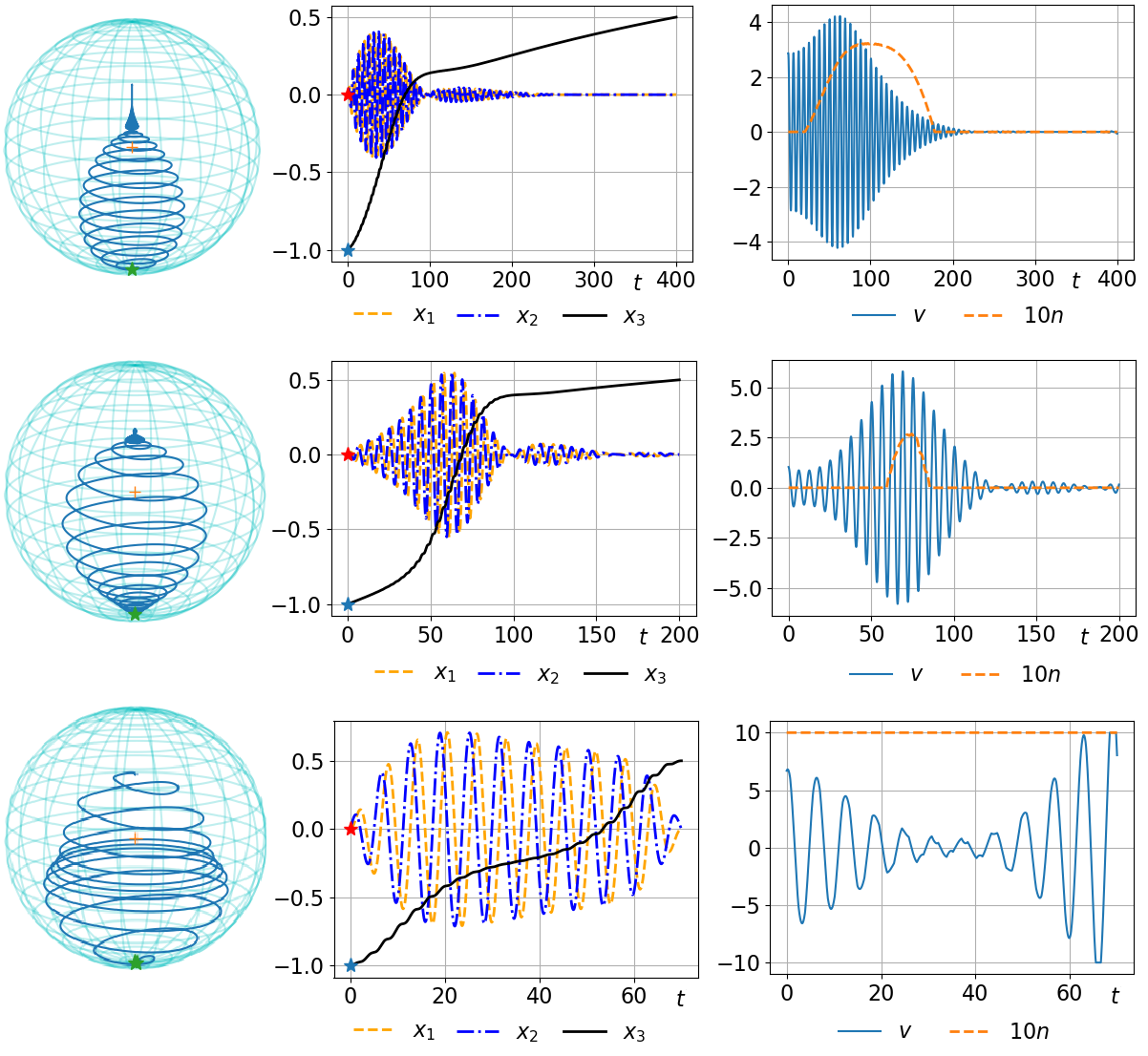} 
	\caption{Moving the system in the Bloch ball from the initial state $(0, 0, -1)$ 
		to the target state $(0, 0, 0.5)$ for $T=400$ (top), $T = 200$ (middle),	and $T = 70$ (bottom).
		Left: evolution of the state in the Bloch ball. Center: evolution of 
		components $x_1$, $x_2$, $x_3$. Right: optimal coherent (solid blue line) and
		incoherent (dash orange line) controls.}
	\label{fig:1}
\end{figure}

\begin{figure}
	\centering
	\includegraphics[scale=0.45]{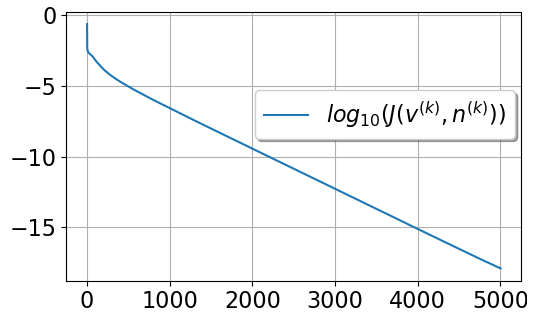} 
	\caption{Decrease of the cost functional $J$ vs iteration number (case $T=70$ which corresponds to the bottom pictures at Figure~\ref{fig:1}).} 
	\label{fig:2}
\end{figure}

Figure~\ref{fig:1} shows the results of numerical optimization 
for moving the system from the initial state with vector
$(0, 0, -1)$ to the target state with vector $(0, 0, 0.5)$ for different values of $T$. 
When $T$ is relatively small, the method gives solution such that
$x_1$, $x_2$, $x_3$ have large amplitudes almost during the entire 
time period. Figure~\ref{fig:2} shows how the cost functional $J$ decreases (in logarithmic scale) vs the iteration number in the sequential updates of $u$ for $T = 70$. 

\begin{figure}
	\centering
	\includegraphics[scale=0.37]{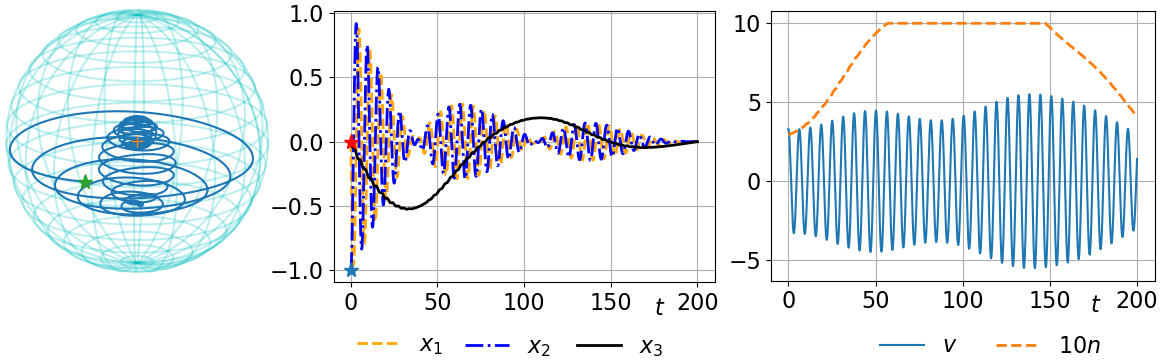} 
	\caption{Moving the system from the initial state $(0, -1, 0)$ to the target state $(0, 0, 0)$.
		Left: evolution of the state in the Bloch ball. Center: evolution of 
		components $x_1$, $x_2$, $x_3$. Right: optimal coherent (solid blue line) and
		incoherent (dash orange line) controls.}
	\label{fig:3}
\end{figure}

Figure~\ref{fig:3} shows the numerical results for moving a pure state with $(0,-1,0)$ into the target completely mixed state $(0,0,0)$. 

\section{Conclusions} 

In this work manipulation of states of a two-level open quantum 
system driven by coherent and incoherent controls is considered. 
The control goal is to steer an initial density matrix into 
a target density matrix in as small as possible final time. 
To achieve this goal, we consider a decreasing series of final 
times, starting from some large enough final time. For each 
final time the control problem is formulated as minimizing 
distance to the target state. Then Pontryagin maximum principle 
for fixed final time and gradient projection method are applied 
to construct a numerical method for solving this problem. Several 
examples of initial and final states are explicitly considered 
using this method and minimal time and corresponding controls 
are numerically found.

\vspace{0.5cm}

{\bf Acknowledgements}. Sections~2,~4~--~6 of this work are performed 
within the Russian Science Foundation Project No.~17-11-01388. 
Derivation of the dynamical equations in Section~3 is performed 
within the project No.~1.669.2016/FPM of the Ministry 
of Science and Higher Education of the Russian Federation.


\newpage

\begin{thebibliography}{9}

\bibitem{RiceBook2000}  
Rice, S.A., Zhao, M.: Optical Control of Molecular Dynamics. Wiley, New York (2000) 

\bibitem{BrumerBook2003}  
Brumer, P.W., Shapiro, M.: Principles of the Quantum Control of Molecular Processes. Wiley-Interscience (2003) 

\bibitem{TannorBook2007}  
Tannor, D.J.: Introduction to Quantum Mechanics: A Time Dependent Perspective.
University Science Press, Sausalito (2007) 

\bibitem{FradkovBook2007} 
Fradkov, A.L.: Cybernetical Physics: From Control of Chaos to Quantum Control. Springer-Verlag, New York (2007)  

\bibitem{Brif_Chakrabarti_Rabitz_article_2010}
Brif, C., Chakrabarti, R., Rabitz, H.: Control of quantum phenomena: past,
present and future. New J. Phys. {\bf 12}(7), 075008 (2010). 
\url{https://doi.org/10.1088/1367-2630/12/7/075008}

\bibitem{WisemanBook2010} 
Wiseman, H.M., Milburn G.J.: Quantum Measurement and Control. 
Cambridge University Press (2010) 

\bibitem{Petersen2010} 
Dong, D., Petersen, I.R.: Quantum control theory and applications: 
a survey. IET Control Theory \& Applications. {\bf 4}(12), 2651--2671 (2010). \url{http://dx.doi.org/10.1049/iet-cta.2009.0508}

\bibitem{ZagoskinBook2011} 
Zagoskin, A.M.: Quantum Engineering: Theory and Design of Quantum 
Coherent Structures. Cambridge University Press, Cambridge (2011) 

\bibitem{Glaser2015Report}  
Glaser, S.J., Boscain, U., Calarco, T., Koch, C.P., 
K\"{o}ckenberger, W., Kosloff, R., Kuprov, I., Luy, B., 
Schirmer, S., Schulte-Herbr\"{u}ggen, T., Sugny, D., 
Wilhelm, F.K.: Training Schr\"{o}dinger’s cat: quantum optimal control.
Strategic report on current status, visions and goals for research in Europe.
Eur. Phys. J. D. {\bf 69}(12), 279 (2015).
\url{https://doi.org/10.1140/epjd/e2015-60464-1}

\bibitem{CPKoch_2016_OpenQS} 
Koch, C.P.: Controlling open quantum systems: Tools, achievements, 
and limitations. J.~Phys.: Condens. Matter. {\bf 28}(21), 213001
(2016). \url{https://doi.org/10.1088/0953-8984/28/21/213001}

\bibitem{Borzi_book_2017}
Borz\`\i~, A., Ciaramella, G., Sprengel, M.: Formulation and Numerical Solution of Quantum Control Problems. SIAM, Philadelphia (2017)

\bibitem{Lyakhov_Pechen_Lee_2018}
Lyakhov, K.A., Pechen, A.N., Lee, H.-J.: The efficiency of one-line versus multi-line excitation of boron
isotopes within the method of selective laser assisted retardation of condensation. AIP Adv. {\bf 8}, 95325
(2018). \url{https://doi.org/10.1063/1.5040903}

\bibitem{Amosov_Mokeev_2018}
Amosov, G.G., Mokeev, A.S.: On non-commutative operator graphs generated by covariant resolutions
of identity. Quantum Inf. Process. {\bf 17}, 325 (2018). \url{https://doi.org/10.1007/s11128-018–2072-x}

\bibitem{Avanesov_Kronberg_Pechen_2018}
Avanesov, A.S., Kronberg, D.A., Pechen, A.N.: Active beam splitting attack applied to differential 
phase shift quantum key distribution protocol. P-Adic Numbers, Ultrametric Analysis and Applications. {\bf 10}(3),
222--232 (2018). \url{https://doi.org/10.1134/S2070046618030068}

\bibitem{Pechen_Rabitz_2006}
Pechen, A., Rabitz, H.: Teaching the environment to control 
quantum systems. Phys. Rev.~A. {\bf 73}, 062102 (2006).
\url{https://doi.org/10.1103/PhysRevA.73.062102}

\bibitem{Pechen_Trushechkin_2015}
Pechen, A., Trushechkin, A.: Measurement-assisted Landau--Zener transitions. Phys. Rev.~A. {\bf 91}(5), 052316 (2015).
\url{https://doi.org/10.1103/PhysRevA.91.052316}

\bibitem{Pechen_Rabitz_2014}
Pechen, A., Rabitz, H.: Incoherent control of open quantum systems.
J.~Math. Sci. {\bf 199}(6), 695--701 (2014). \url{https://doi.org/10.1007/s10958-014-1895-y}

\bibitem{Pechen_PhysRevA_2011}
Pechen, A.: Engineering arbitrary pure and mixed quantum states.
Phys. Rev.~A. {\bf 84}, 042106 (2011). \url{https://doi.org/10.1103/PhysRevA.84.042106}

\bibitem{Lapert_2010} 
Lapert, M., Zhang, Y., Braun, M., Glaser, S.J., Sugny, D.: Singular extremals for the time-optimal control of dissipative spin 1/2 particles. Phys. Rev. Lett. {\bf 104}, 083001 (2010). \url{https://doi.org/10.1103/PhysRevLett.104.083001}

\bibitem{Boscain_Gronberg_Long_Rabitz_2014}
Boscain, U., Gr\"{o}nberg, F., Long, R., Rabitz, H.: Minimal time trajectories for two-level quantum systems with two bounded controls. J.~Math. Phys. {\bf 55}, 062106 (2014). \url{https://doi.org/10.1063/1.4882158}

\bibitem{Albertini_DAlessandro_2016}
Albertini, F., D'Alessandro, D.: Time optimal simultaneous control 
of two level quantum systems. Automatica. {\bf 74}, 55--62 (2016).
\url{https://doi.org/10.1016/j.automatica.2016.07.014} 

\bibitem{Caneva2009} 
Caneva, T., Murphy, M., Calarco, T., Fazio, R., Montangero, S., 
Giovannetti, V., Santoro G.E.: Optimal control at the quantum speed limit. Phys. Rev. Lett. {\bf 103}, 240501 (2009). \url{https://doi.org/10.1103/PhysRevLett.103.240501}

\bibitem{Volovich_Kozyrev_2016}
Volovich, I.V., Kozyrev, S.V.: Manipulation of states of a degenerate quantum system.
Proc. Steklov Inst. Math. {\bf 294}, 241--251 (2016).
\url{https://doi.org/10.1134/S008154381606016X}

\bibitem{Pontryagin_et_al_book_1962}
Pontryagin, L.S., Boltyanskii, V.G., Gamkrelidze, R.V., Mishchenko, E.F.: The Mathematical Theory of Optimal Processes / Transl. from Russian. John Wiley \& Sons, Inc., New York, London (1962)

\bibitem{Nikolskii_2007}
Nikol’skii, M.S.: Convergence of the gradient projection method in optimal control problems. Comput. Math. Model. {\bf 18}(2), 148--156 (2007). \url{https://doi.org/10.1007/s10598-007-0015-y}

\bibitem{Demyanov_Rubinov_book_1970}
Demyanov, V.F., Rubinov, A.M.: Approximate methods in optimization problems / Transl. from Russian. American Elsevier Pub. Co., New York (1970) 

\bibitem{Bertsekas_book_2016}
Bertsekas, D.P.: Nonlinear Programming. 3rd ed. Athena Scientific, Belmont, MA (2016)

\bibitem{Holevo_book_De_Gruyter_2012}
Holevo, A.S.: Quantum Systems, Channels, Information.
Walter de Gruyter GmbH, Berlin, Boston (2012)  
(Ser.: De Gruyter Studies in Mathematical Physics, 16). 

\bibitem{Sklarz_Tannor_article_2002}
Sklarz, S.E., Tannor, D.J.: Loading a Bose-Einstein condensate onto
an optical lattice: An application of optimal control theory to the nonlinear Schr\"{o}dinger equation. Phys. Rev.~A. {\bf 66}(5), 053619 (2002). \url{https://doi.org/10.1103/PhysRevA.66.053619} 

\bibitem{Khaneja_Reiss_Kehlet_SchulteHerbruggen_Glaser_2005}
Khaneja, N., Reiss, T., Kehlet, C., Schulte-Herbr\"{u}ggen, T., Glaser, S.J.: Optimal control of coupled spin dynamics: design of NMR pulse sequences by gradient ascent algorithms. J.~Magn.~Reson. {\bf 172}(2), 296--305, (2005).  \url{https://doi.org/10.1016/j.jmr.2004.11.004}

\bibitem{Caneva_Calarco_Montangero_2011}
Caneva, T., Calarco, T., Montangero, S.: Chopped random-basis 
quantum optimization. Phys. Rev.~A. {\bf 84}(2), 022326 (2011).
\url{https://doi.org/10.1103/PhysRevA.84.022326}

\bibitem{Ilin_Shpagina_Uskov_Lychkovskiy_article_2018}
Il'in, N., Shpagina, E., Uskov, F., Lychkovskiy, O.: 
Squaring parametrization of constrained and unconstrained 
sets of quantum states. J.~Phys.~A. {\bf 51}(8), 85301 (2018).
\url{https://doi.org/10.1088/1751-8121/aaa32d}	 
	 
\end{thebibliography}
\end{document}